# ENERGY DEPOSITION AND RADIOLOGICAL STUDIES FOR THE LBNF HADRON ABSORBER[*][§]

I.L. Rakhno[#], N.V. Mokhov, I.S. Tropin

Fermilab, Batavia, IL 60510, USA

Y.I. Eidelman

Euclid TechLabs, LLC, Solon, OH 44139, USA

## Abstract

Results of detailed Monte Carlo energy deposition and radiological studies performed for the LBNF hadron absorber with the MARS15 code are described. The model of the entire facility, that includes a pion-production target, focusing horns, target chase, decay channel, hadron absorber system – all with corresponding radiation shielding – was developed using the recently implemented ROOT-based geometry option in the MARS15 code. Both normal operation and accidental conditions were studied. Results of detailed thermal calculations with the ANSYS code helped to select the most viable design options.

[*]Work supported by Fermi Research Alliance, LLC, under contract No. DE-AC02-07CH11359 with the U.S. Department of Energy.
[§]Presented at the 6th International Particle Accelerator Conference (IPAC´15), May 3-8, 2015, Richmond, VA, USA.
[#]rakhno@fnal.gov

# ENERGY DEPOSITION AND RADIOLOGICAL STUDIES FOR THE LBNF HADRON ABSORBER*

I.L. Rakhno#, N.V. Mokhov, I.S. Tropin, Fermilab, Batavia, IL 60510, USA
Y.I. Eidelman, Euclid TechLabs, LLC, Solon, OH 44139, USA

*Abstract*

Results of detailed Monte Carlo energy deposition and radiological studies performed for the LBNF hadron absorber [1] with the MARS15 code [2] are described. The model of the entire facility, that includes a pion-production target, focusing horns, target chase, decay channel, hadron absorber system – all with corresponding radiation shielding – was developed using the recently implemented ROOT-based geometry option in the MARS15 code. Both normal operation and accidental conditions were studied. Results of detailed thermal calculations with the ANSYS code helped to select the most viable design options.

## INTRODUCTION

The Long-Baseline Neutrino Facility (LBNF) at Fermilab is supposed to provide the world's highest-intensity neutrino beam for the US and international programs in neutrino physics [1]. The corresponding incoming proton beam power can ultimately be as high as 2.4 MW, and the underground beam absorber at the end of the decay channel with related infrastructure is supposed to operate with little or no maintenance for about 20 years. Such a combination of long operation time and high deposited power imposes strict limitations on design.

## UNIFIED COMPUTER MODEL

A very detailed model of the entire facility which includes a pion-production target, focusing horns, target chase, decay channel, and hadron absorber system was developed. A drawing that describes an elevation view of the entire facility and a fragment of the entire MARS model which shows the absorber hall are presented in Fig. 1. In the Figure, light blue and gray colors refer to air and concrete, respectively, and green color means soil. As a result of thorough optimization studies with MARS code, the system now consists of a spoiler aluminum block, five aluminum mask blocks, nine sculpted aluminum blocks, four solid aluminum blocks, and four central steel blocks, and all that is surrounded with sophisticated steel and concrete shielding. Total weight of the aluminum and steel is 39 and 2,500 ton, respectively. Volume of poured concrete is 24,000 ft$^3$.

## BEAM PARAMETERS AND SCENARIOS

Table 1 summarizes the beam parameters used to study both the normal operation and two most severe credible accident scenarios. The beam starts at z=-7.3 m (this location is labelled as MC0 in Table 1) and emittance $\varepsilon_{95}$ is 20$\pi$ mm-mrad. The tilt downward is 0.101074 radian.

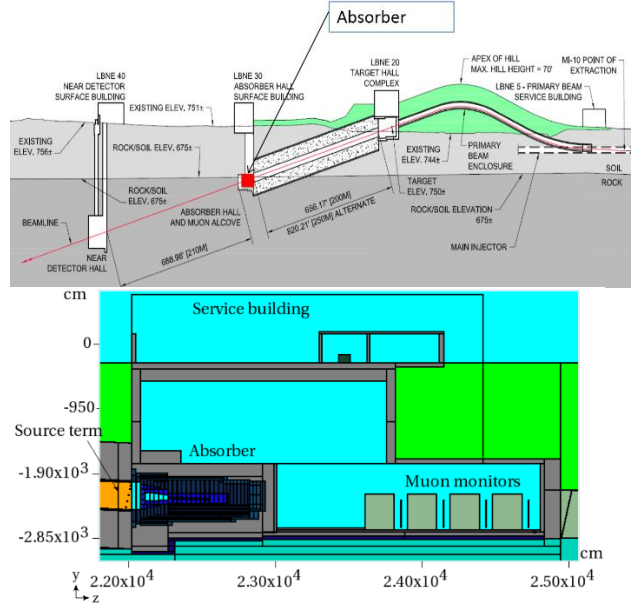

Figure 1: Elevation view of the entire facility (top) and a fragment of the MARS model that shows the hadron absorber and muon monitors in the hall with a service building above (bottom).

Table 1: Beam Parameters and Studied Scenarios

| Parameter | Normal Operation | No-Target Accident On-Axis | Off-Axis Accident§ |
|---|---|---|---|
| $E_p$(GeV) | 120<br>60 | 120 | 120 |
| P(MW) or Q(MJ) | 2.40 MW<br>2.06 MW | 2.88 MJ | 2.88 MJ |
| $\sigma_0$ (mm) at MC0 | 1.7<br>1.7 | 2.4 | 2.4 |
| $\beta_0$ (m) at MC0 | 110.8837<br>55.44 | 221.03 | 221.03 |
| Cycle (s) | 1.2<br>0.7 | 1.2 | 1.2 |
| Intensity ($10^{14}$) | 1.25 p/s<br>2.14 p/s | 1.5 p/pulse | 1.5 p/pulse |

§Beam points to absorber cooling water pipes.

*Work supported by Fermi Research Alliance, LLC, under contract No. DE-AC02-07CH11359 with the U.S. Department of Energy
#rakhno@fnal.gov

## INCOMING SOURCE TERM

When calculating the incoming source term, interactions in both the target and decay channel were taken into account, so that it was possible to study the effect of replacement of the air with helium in the decay pipe. The hadron absorber design is more challenging for the case of helium-filled decay pipe because of higher peak energy deposition. The calculated distributions of the source term across the beam pipe cross section are shown in Fig. 2.

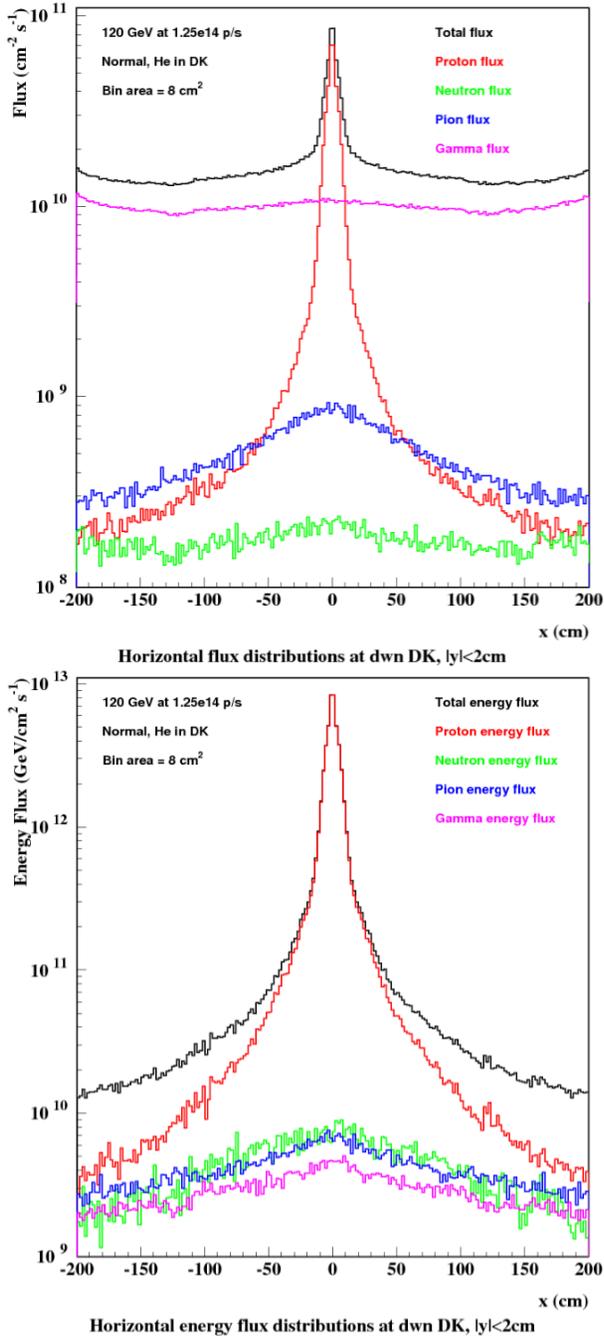

Figure 2: Incoming particle (top) and energy (bottom) flux at absorber for normal operation at 120 GeV with helium in the decay pipe.

## ENERGY DEPOSITION AND DYNAMIC HEAT LOAD

Driven by the need to reduce the peak temperature in aluminum core to below 100 C at normal operation, a lot of energy deposition optimization studies were performed followed by corresponding thermal and stress analyses with ANSYS code. As a result of these studies, the following design for the absorber core was selected: aluminum (6061-T6) spoiler followed by five aluminum mask blocks, nine sculpted aluminum blocks, four solid aluminum blocks, and four central steel blocks. Energy deposition isocontours are given in Fig. 3. Total power deposited in the absorber and surrounding shielding at normal operation at 120 GeV is 742 kW, and power deposited in aluminum, steel, concrete and other components is 501, 235, 2.4, and 3.5 kW, respectively.

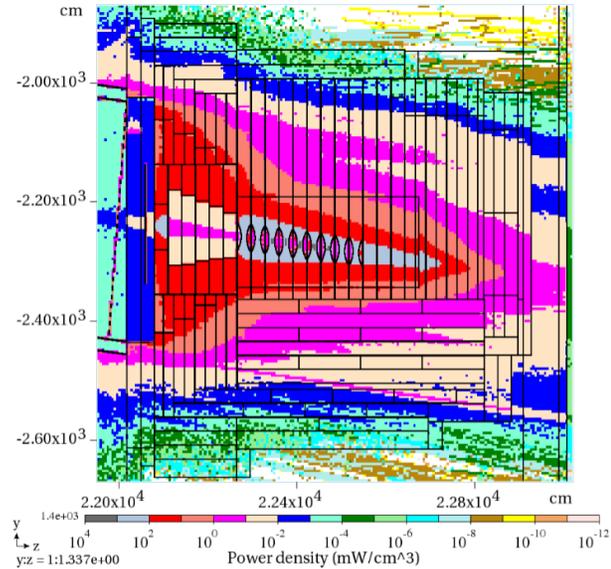

Figure 3: Calculated energy deposition distribution at 120 GeV for normal operation.

## PEAK VALUES AND RADIATION DAMAGE IN THE CORE

Calculated peak values in the absorber core and radiation damage for normal operation at 120 GeV are provided in Table 2. One can see that radiation damage is not an issue in this case due to a substantial predicted safety factor.

## RADIOLOGICAL RESULTS

### Prompt and Residual Dose

According to existing DOE regulations, prompt dose in all areas accessible by general public should not exceed 0.5 μSv/hr, while in areas of unlimited access for radiation workers the dose should not exceed 2.5 μSv/hr. Residual dose—after 100-day irradiation and 4-hour cooling—in areas accessible by radiation workers (beam-off) should not exceed 50 μSv/hr. The calculated distributions of prompt and residual dose around the facility are shown in Fig. 4. One can see that the predicted values comply with the regulatory requirements.

Table 2: Peak Values in Aluminum and Steel in the Absorber Core. The hadron flux, F, is above 0.1 MeV

| Quantity | Al Spoiler | Al Sculpted #3 | Al Solid #1 | Steel #1 |
|---|---|---|---|---|
| PD (W/cm$^3$) | 1.8 | 1.8 | 0.77 | 0.78 |
| DPA per 20 years | 0.62 | 0.48 | 0.20 | 0.34 |
| DPA$_{limit}$* | 5-10 | 5-10 | 5-10 | 10 |
| F(cm$^{-2}$) per 20 years, $10^{20}$ | 4.48 | 4.41 | 1.69 | 1.88 |
| F$_{limit}$*(cm$^{-2}$), $10^{20}$ | 50 | 50 | 50 | 700 |

*Limit values: No significant effect on specific heat, swelling, elongation, elastic and tensile properties up to the values shown.

## Groundwater Activation and Air Releases

Groundwater activation calculated according to Fermilab Concentration model [3] is well—with a safety factor of ten—below the limit. The calculation deals with maximum hadron flux in the surrounding soil and rock, and for this model the maximum hadron flux was found to be 400 cm$^{-2}$ s$^{-1}$.

Air releases from the absorber system have been considered by K. Vaziri using hadron fluxes above 30 MeV calculated with MARS code in air pockets inside absorber and in various regions of the absorber hall and muon alcove [4]. Three cases were studied for a 120-GeV beam at 2.4 MW beam power:

- Activated air from absorber core is directly sent to the air handling room and then to the target hall.
- Air from the absorber hall, with no contribution from the core air, is sent to the air handling room and then to the target hall.
- Air from the absorber core is released to the absorber hall and the combined air is sent to the air handling room and then to the target hall.

The third case is the most realistic air release scenario. In this case, a conservative calculation results in 7.57 Ci of absorber hall activated air to be released from the target hall annually. The air released from the target hall will be sent through further decay volumes, hence the absorber hall released activity will be further reduced. As a result, the activated air released from the current configuration of the LBNF absorber will conservatively contribute less than 2% to the total activated air release budget of the laboratory.

In conclusion, the shielding configuration brings the prompt and residual dose as well as groundwater and air activation levels to well below the Fermilab regulatory limits for radiation workers and general public.

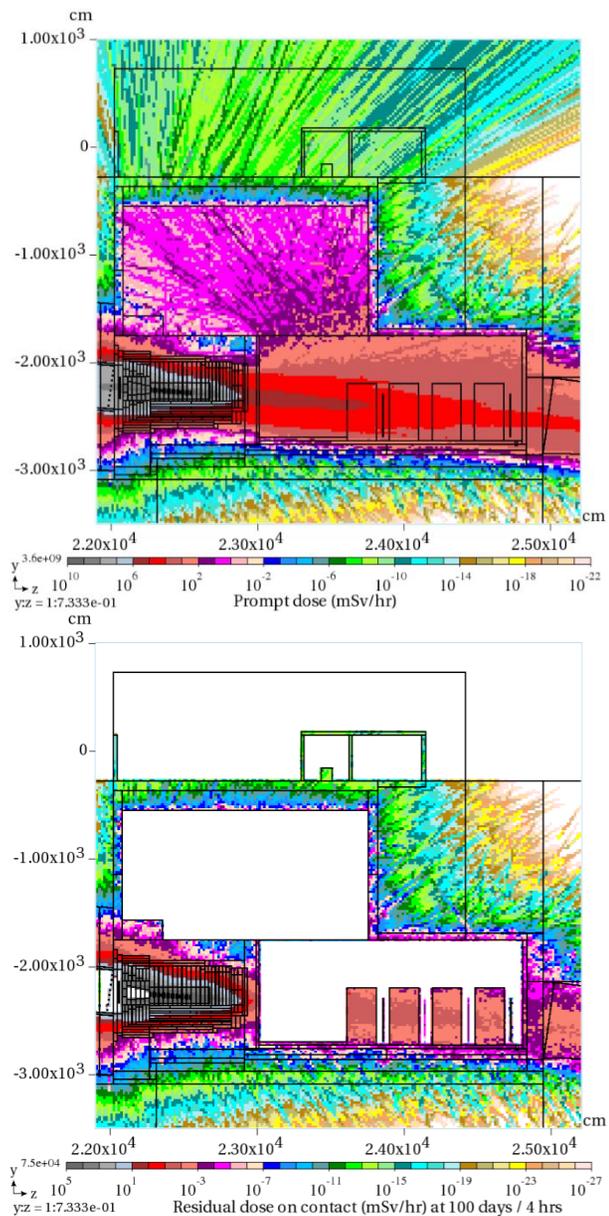

Figure 4: The calculated distributions of prompt (top) and residual (bottom) dose around the facility at normal operation at 120 GeV.